\documentclass[useAMS,usenatbib]{mn2e}

\usepackage{aecompl}

\usepackage{graphicx}
\usepackage{amssymb}
\usepackage{amsmath}
\usepackage{aas_macros}
\usepackage{deluxetable}
\usepackage{lscape}
\usepackage{rotating}
\usepackage[symbol*]{footmisc}



\title[Perturbation growth in accreting filaments]{Perturbation growth in accreting filaments}
\author[S. D. Clarke, A. P. Whitworth and D. A. Hubber]{S. D. Clarke$^{1}$\thanks{E-mail: seamus.clarke@astro.cf.ac.uk }, A. P. Whitworth$^{1}$ and D. A. Hubber$^{2}$ $^{3}$\\$^{1}$School of Physics and Astronomy, Cardiff University, Cardiff, CF24 3AA, UK\\$^{2}$University Observatory Munich, Ludwig-Maximilians-University Munich, Scheinerstr.1, D-81679 Munich, Germany\\$^{3}$Excellence Cluster Universe, Boltzmannstr. 2, D-85748 Garching, Germany}

\newcommand{\OO}{_{_{\rm O}}}
\newcommand{\Su}{_{_{\odot}}}
\begin{document}

\date{}

\pagerange{\pageref{firstpage}--\pageref{lastpage}} \pubyear{2002}

\maketitle

\label{firstpage}

\begin{abstract}

We use smoothed particle hydrodynamic simulations to investigate the growth of perturbations in infinitely long filaments as they form and grow by accretion. The growth of these perturbations leads to filament fragmentation and the formation of cores. Most previous work on this subject has been confined to the growth and fragmentation of equilibrium filaments and has found that there exists a preferential fragmentation length scale which is roughly 4 times the filament's diameter. Our results show a more complicated dispersion relation with a series of peaks linking perturbation wavelength and growth rate. These are due to gravo-acoustic oscillations along the longitudinal axis during the sub-critical phase of growth. The positions of the peaks in growth rate have a strong dependence on both the mass accretion rate onto the filament and the temperature of the gas. When seeded with a multi-wavelength density power spectrum there exists a clear preferred core separation equal to the largest peak in the dispersion relation. Our results allow one to estimate a minimum age for a filament which is breaking up into regularly spaced fragments, as well as an average accretion rate. We apply the model to observations of filaments in Taurus by \citet{TafHac15} and find accretion rates consistent with those estimated by \citet{Pal13}.     

\end{abstract}


\section{Introduction}%

The prevalence of filamentary structures across a wide range of scales \citep{SchElm79,Lad99,Mye09,And10,Pal13,Beu15} has lead to several papers studying their structure, stability, fragmentation and collapse \citep{Ost64,Lar85,InuMiy92,BurHar04,Pon11,FisMar12,Hei13a,ClaWhi15}. 

It has been shown that a filament's line density (defined as the mass per unit length) determines the filament's radial stability: filaments below a critical line density will not and cannot be made to collapse radially, those with line density above the critical value will. For an isothermal filament the critical line density is 
\begin{equation}
\mu_{_{\rm{CRIT}}} \; = \; \frac{2a_{\rm{o}}^{2}}{G} \; \approx \; 16.7 \, \rm{M\Su} \, \rm{pc}^{-1} \; \left( \frac{T}{10 \, \rm{K}} \right),
\end{equation}
where $a_{\rm{o}}$ is the isothermal sound speed, $G$ is the gravitational constant and $T$ is the gas temperature \citep{Ost64}. 

The Herschel Gould Belt Survey has shown that the majority of star-forming cores are found within filaments which have super-critical line densities, while sub-critical filaments are sterile \citep{And10}. It has also been shown theoretically that due to their geometry, filaments are apt to fragment; small-scale perturbations can readily collapse locally before global longitudinal collapse overwhelms them \citep{Pon11}. This suggest a paradigm in which filaments are formed inside molecular clouds, and the densest of these filaments then become super-critical and go on to fragment into cores. 

\citet{InuMiy92} have analysed how small-scale perturbations grow and collapse in an equilibrium filament; they derive a dispersion relation linking perturbation wavelength with perturbation growth rate. They find that perturbations are unstable when their wavelength is larger than twice the filament's diameter, and there exists a fastest growing mode at approximately 4 times the filament's diameter. When the line density of an isothermal filament exceeds the critical value by a small amount, the perturbations do not have time to grow before global radial collapse takes place; in this case, it is thought that fragmentation occurs at the point when isothermality breaks and radial collapse is halted.

Though perturbations in equilibrium and super-critical filaments have been studied before \citep[e.g.][]{Lar85,InuMiy92,InuMiy97,Fre14}, non-equilibrium filaments have been neglected. However, it is unlikely that when a filament first forms it is in equilibrium or has a super-critical line density; it is far more likely that a filament will be sub-critical when it first forms and mass then accretes on to it until it becomes unstable and fragments. As filaments and perturbations form together and co-evolve it is important to understand how density perturbations behave during the sub-critical phase, especially so if one attempts to link the density perturbation power spectrum to the core mass function \citep{Inu01,Roy15}. 

In this paper, we present numerical simulations of initially sub-critical perturbed accreting filaments, in order to investigate the dispersion relation between perturbation wavelength and perturbation growth rate, and to compare to the relation found by \citet{InuMiy92} for equilibrium filaments. In section \ref{SEC:NUM}, we detail the simulation setup and the initial conditions used; in section \ref{SEC:RES}, we present the results of these simulations; in section \ref{SEC:DIS}, we discuss their significance and compare to previous work; and in section \ref{SEC:CON}, we summarize our conclusions.   

\section{Numerical Setup}\label{SEC:NUM}%

\begin{figure}
\centering
\includegraphics[width = 0.98\linewidth]{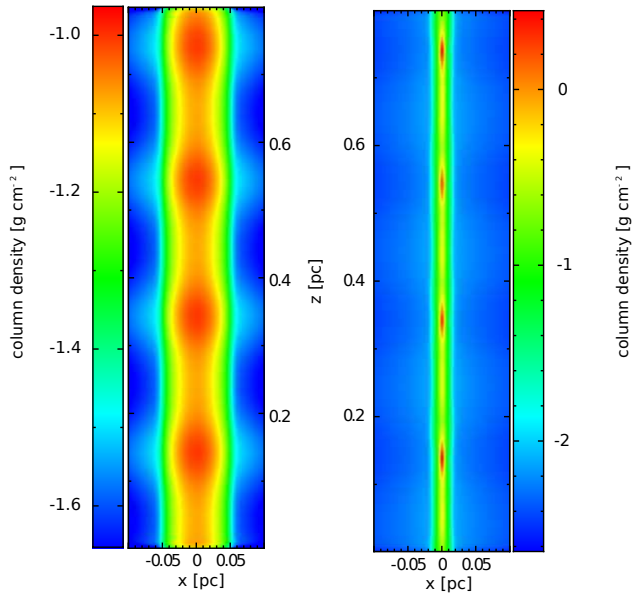}
\caption{The column density projected onto the $x$-$z$ plane for the fiducial case (i.e. $A=0.2$, $\lambda = 0.2 \, \rm pc$, $T\OO = 40 \, \rm K$ and $\dot{\mu} = 70 \, \rm{M\Su \, pc^{-1} \, Myr^{-1}}$) at two different times. On the left, at $t = 0.15 \, \rm Myr$, the filament has formed on the $z$ axis, it is sub-critical and confined by the ram pressure of the accreting gas. On the right, at $t = \, \rm 0.55 Myr$, the filament has become supercritical and is contracting radially; at the same time the seeded perturbations have become sites of local collapse. Created using the visualisation tool SPLASH \citep{Pri07}.}
\label{fig::fil}
\end{figure} 

The simulations presented in this paper are performed using the Smoothed Particle Hydrodynamics (SPH) code GANDALF (Hubber et al. in prep). The simulations invoke both self-gravity and hydrodynamics, with the barotropic equation of state
\begin{equation}
T(\rho) = T\OO\left(1 + \left( \frac{\rho}{\rho_{_{\rm BARO}}} \right)^{2/3} \right).
\end{equation} 
Here $T\OO$ is the initial temperature and $\rho_{_{\rm BARO}} = 10^{-14} \; \rm{g \, cm^{-3}}$ is the critical density at which the gas changes from being approximately isothermal to approximately adiabatic. We present two sets of simulations, one with $T\OO  =  10 \, \rm{K}$ and the other with $T\OO = 40 \, \rm{K}$; this results in isothermal sound speeds of $\sim \, 0.19 \, \rm{km/s}$ and $\sim \, 0.37 \, \rm{km/s}$ respectively, assuming solar metallicity. Grad-h SPH \citep{PriMon04} is implemented, with $\eta=1.2$, so that a typical particle has $\sim 56$ neighbours. Sink particles are implemented as described in \citet{Hub13} using the sink creation density, $\rho_{_{\rm SINK}} = 10^{-12} \, \rm{g \, cm^{-3}}$.

The computational domain is open in the $x$ and $y$ directions, but is periodic in $z$, the long axis of the filament (see Wunsch et al. in prep, for the implementation of the modified Ewald field). This in effect allows us to study the perturbation growth in an infinitely long filament, and hence to ignore the complicating effects of global longitudinal collapse \citep{ClaWhi15}.

To generate the initial conditions, a cylindrical settled glass of particles with a uniform density is stretched so that it reproduces the density profile,
\begin{equation}
\rho(r,z) \, = \, \frac{\rho\OO r\OO}{r} \left(1 + A \sin{\left(\frac{2 \pi z}{\lambda}\right)}\right).
\end{equation}  
Here $\rho\OO$ is the density at $r\OO = 0.1 \, \rm{pc}$, $A$ is the amplitude of the perturbation and $\lambda$ is the perturbation's wavelength. The initial velocity field is,

\begin{equation}
\underline{v} = -v\OO \underline{\hat{r}}.
\end{equation} 
Combined, these density and velocity profiles result in a setup which can be characterised by $A$, $\lambda$ and the influx of mass per unit length, $\dot{\mu} = 2\pi\rho\OO r\OO v\OO$.

The resulting perturbed cylinder of particles has a radius of $r_{max} = 1 \, \rm{pc}$, and a length, $L = m \lambda$, where $m$ is the largest integer that satisfies $L \leq 1 \, \rm{pc}$. This provides a sufficiently large computational domain to allow us to study a wide range of plausible perturbation wavelengths, while maintaining good resolution.  

We stress that there is no filament present at the outset. At $t = 0$ all the material is flowing radially inwards, towards the $z$-axis. As soon as the simulation starts, an accretion shock forms on the $z$-axis and then propagates outwards; the filament is the dense material inside this shock. Our initial setup should be seen as a simple approximation to the formation of a filament, a locally convergent flow in a globally turbulent field, which is about to create a filament. 

We consider perturbations which are initially small, taking $A = 0.1$ or $0.2$. The perturbation wavelength $\lambda$ is varied between $0.05 \, \rm pc$ and $0.5 \, \rm pc$. The lower limit is due to resolution concerns, and the upper limit to ensure that at least 2 wavelengths fit within the computational domain. The mass flux per unit length is varied between $\sim 10 \, \rm{M\Su \, pc^{-1} \, Myr^{-1}}$ and $\sim 100 \, \rm{M\Su \, pc^{-1} \, Myr^{-1}}$, in line with the accretion rates estimated observationally by \citet{Pal13} and \citet{Kir13}. We define the critical time as the time at which the filament reaches the critical line density and becomes radially unstable, $\tau_{_{\rm CRIT}} \sim \mu_{_{\rm CRIT}} / \dot{\mu}$. 

After the first sink forms, we only follow the simulation for a further 0.01 Myr, since we are principally interested in the dynamics that lead to instability, rather than the subsequent collapse. 

The simulations are run with 3 million particles per parsec, giving a mass resolution of between $1.5 \, \times \, 10^{-3} \, \rm{M\Su}$ and $4 \, \times \, 10^{-3} \, \rm{M\Su}$. Such high resolution is necessary to ensure the small wavelength perturbations are well resolved even at low densities. The artificial viscosity needed to capture shocks is known to overly dampen short wavelength oscillations on the order of the smoothing length. Therefore, we have invoked time dependent artificial viscosity as described in \citet{MorMon97} to further lessen the effect of artificial viscosity on short wavelength oscillations. Tests with different numbers of particles show that the simulations have converged at 3 million particles per parsec.

\section{Results}\label{SEC:RES}%

\begin{figure}
\centering
\includegraphics[width = 0.98\linewidth]{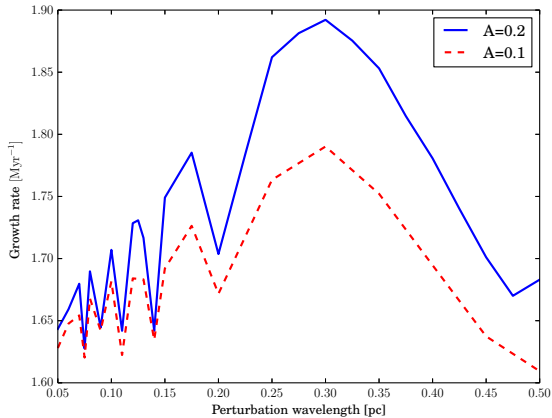}
\caption{The dispersion relation between perturbation wavelength and perturbation growth rate for $T\OO = 40\, \rm K$, $\dot{\mu} = 70 \, \rm{M\Su \, pc^{-1} \, Myr^{-1}}$ and $A=0.2$ (the fiducial case) and also with $A=0.1$. There no longer exists a single local maximum at 4 times the filament's diameter, instead there exists a series of peaks and troughs. The initial amplitude of the perturbations does not affect the shape of the dispersion relation, rather it only affects the magnitude of the growth rate.}
\label{fig::dis}
\end{figure}

\begin{figure}
\centering
\includegraphics[width=0.98\linewidth]{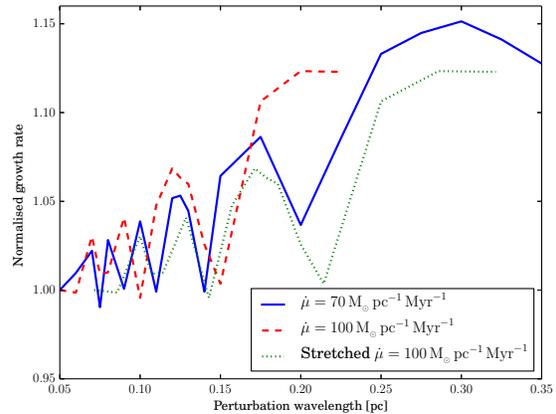}
\caption{The dispersion relation between perturbation wavelength and normalised growth rate for the fiducial case ($\dot{\mu} = 70 \, \rm{M\Su \, pc^{-1} \, Myr^{-1}}$), and for the case with $\dot{\mu} = 100 \, \rm{M\Su \, pc^{-1} \, Myr^{-1}}$. The normalised growth rate is defined as $\hat{g}_{\lambda} = g_{\lambda} / g_{_{0.05 \rm pc}}$. The fiducial case (solid blue line) and the case with a higher accretion rate (dashed red line) are out of phase, the peaks in the dispersion relation do not line up. The higher accretion rate has caused the relation to be squeezed in the x-direction. The green dotted line is the result of stretching the high accretion line assuming the dispersion relation takes the form $g_{\lambda} = f(\lambda/\tau_{_{\rm CRIT}})$, where $\tau_{_{\rm CRIT}}$ is the time at which the filament becomes supercritical. The peaks in the stretched dispersion relation now line up with those from the fiducial case.}
\label{fig::dismu}
\end{figure}

\begin{figure}
\centering
\includegraphics[width=0.98\linewidth]{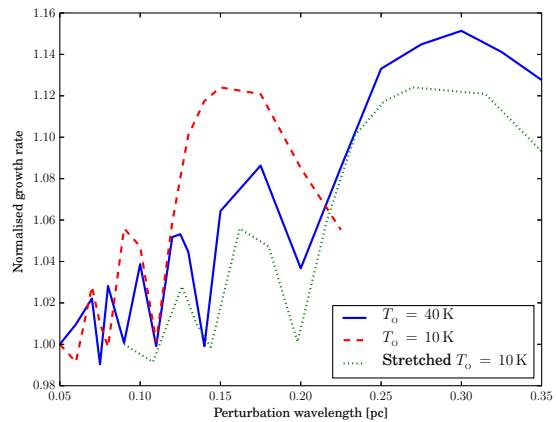}
\caption{The dispersion relation between perturbation wavelength and normalised growth rate for the fiducial case ($T\OO = 40\, \rm K$), and for the case with $T\OO = 10\, \rm K$. The normalised growth rate is defined as $\hat{g}_{\lambda} = g_{\lambda} / g_{_{0.05 \rm pc}}$. The fiducial case (solid blue line) and the case with the lower temperature (dashed red line) are out of phase. The lower temperature has caused the relation to be squeezed in the x-direction. The green dotted line is the result of stretching the $T\OO = 10\, \rm K$ line assuming the dispersion relation takes the form $g_{\lambda} = f'(\lambda  / a\OO)$, where $a\OO$ is the isothermal sound speed. The peaks in the stretched dispersion relation now line up with those from the fiducial case.}
\label{fig::disT}
\end{figure}

The perturbations seeded in the initial density profile survive the approximately cylindrical accretion shock bounding the filament, and form a radially sub-critical perturbed filament of typical width $\sim 0.1 \, \rm{pc}$ (figure \ref{fig::fil}a). The rest of the gas continues to accrete onto the filament until it reaches the critical line density and the perturbed sections collapse (figure \ref{fig::fil}b).

We take the time at which the first sink particle forms, $\tau_{_{\rm SINK}}$ , as a proxy for the perturbation growth time. Sink particles are used in numerical simulations to reduce computational cost; they are used to replace dense, collapsing bound regions that will inevitably become stars. Thus the earlier a sink particle forms the faster the perturbation became unstable. Specifically we define the perturbation growth rate as $g_{\lambda} = 1 / \tau_{_{\rm{SINK}}}$. Figure \ref{fig::dis} shows the dispersion relation linking perturbation wavelength with perturbation growth rate for $T\OO = 40\, \rm K$, $\dot{\mu} = 70 \, \rm{M\Su \, pc^{-1} \, Myr^{-1}}$ and $A=0.2$ (the fiducial case) and also with $A=0.1$. It is evident that the inclusion of the non-equilibrium sub-critical phase has dramatically changed the relationship from the one derived by \citet{InuMiy92}. There no longer exists a single local maximum at 4 times the filament's diameter. Instead the dispersion relation appears to have two features: longer wavelength perturbations tend to grow faster, and superimposed on this there is a series of peaks and troughs.

Varying the mass accretion rate we find that the shape of the dispersion relation is unchanged, but it is rescaled in the x-direction (figure \ref{fig::dismu}). As the mass accretion rate is increased, and the time taken for the filament to become supercritical is decreased, the dispersion relation is squeezed and the peaks in the growth rate move to shorter wavelengths. Let us consider that there exists a function, f(x), which transforms the parameters, $\lambda$, $a\OO$ and $\dot{\mu}$, into the observed dispersion relation $g_{\lambda}$. Figure \ref{fig::dismu} shows that the dispersion relation takes the form $g_{\lambda} = f(\lambda \dot{\mu}) = f(\lambda/\tau_{_{\rm CRIT}})$, where $\tau_{_{\rm CRIT}}$ is the time at which the filament becomes supercritical.

\begin{figure}
\centering
\includegraphics[width=0.98\linewidth]{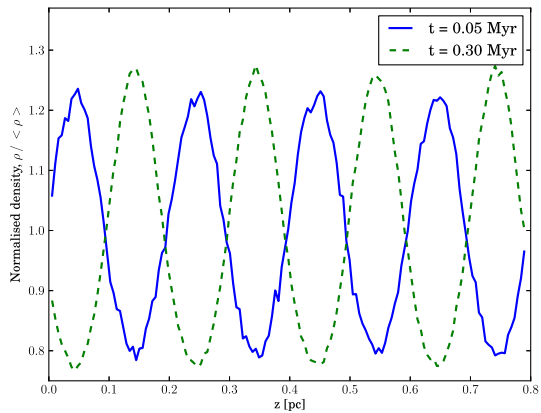}
\caption{The normalised volume-density profile, $\rho \, /\left< \rho \right>$, along the $z$-axis for the fiducial case at two different times, $t = 0.05 \, \rm Myr$ (blue solid line) and $t = 0.30 \, \rm Myr$ (green dashed line). A standing gravo-acoustic wave is set up along the filament's length; the locations of the initial density peaks are the anti-nodes of the wave. The density peaks and trough switch after 0.25 Myr has passed, i.e. half an oscillation period.}
\label{fig::den}
\end{figure}

\begin{figure}
\centering
\includegraphics[width = 0.98\linewidth]{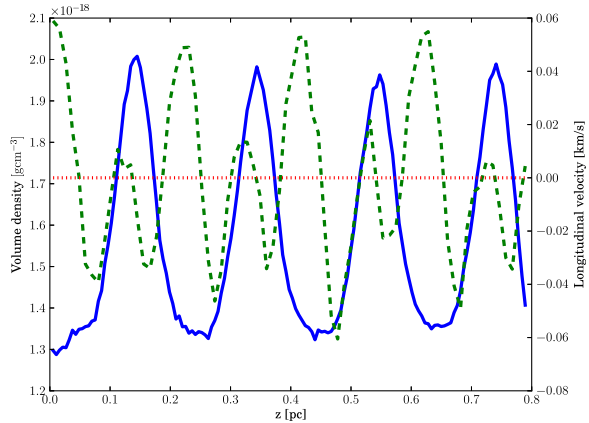}
\caption{The longitudinal density (blue solid line) and velocity (green dashed line) profiles at $t = 0.45 \, \rm{Myr}$ for the simulation with $\lambda = 0.2 \, \rm pc$, $T\OO = 40\, \rm K$, $\dot{\mu} = 70 \, \rm{M\Su \, pc^{-1} \, Myr^{-1}}$ and $A=0.2$. This is the time at which the perturbations are just becoming self-gravitating and the filament is close to becoming globally super-critical. The red dotted line is the $ v = 0 \, \rm{km / s}$ line, which we include to help the reader see the converging and diverging regions along the filament. The velocity field is out of phase with the density field, and the main converging flows are positioned where the troughs of the density profile are. Because of this the majority of the gas is moving away from the density peaks just as collapse is about to begin. There are small converging flows at the central density peaks because these peaks have just become self-gravitating.}
\label{fig::vel}
\end{figure}

When the initial temperature of the gas, and by extension the isothermal sound speed, is varied we see the same type of behaviour as when the mass accretion rate is varied (figure \ref{fig::disT}). As the temperature of the gas is decreased the dispersion relation is squeezed and the peaks in the growth rate move to shorter wavelengths. We find that the stretching goes as $T\OO^{-1/2}$ or as $a\OO^{-1}$, so we can write the dispersion relation as $g_{\lambda} = F(\lambda /a\OO \tau_{_{\rm CRIT}})$.

\section{Discussion}\label{SEC:DIS}%

\begin{figure*}
\centering
\includegraphics[width=0.49\linewidth]{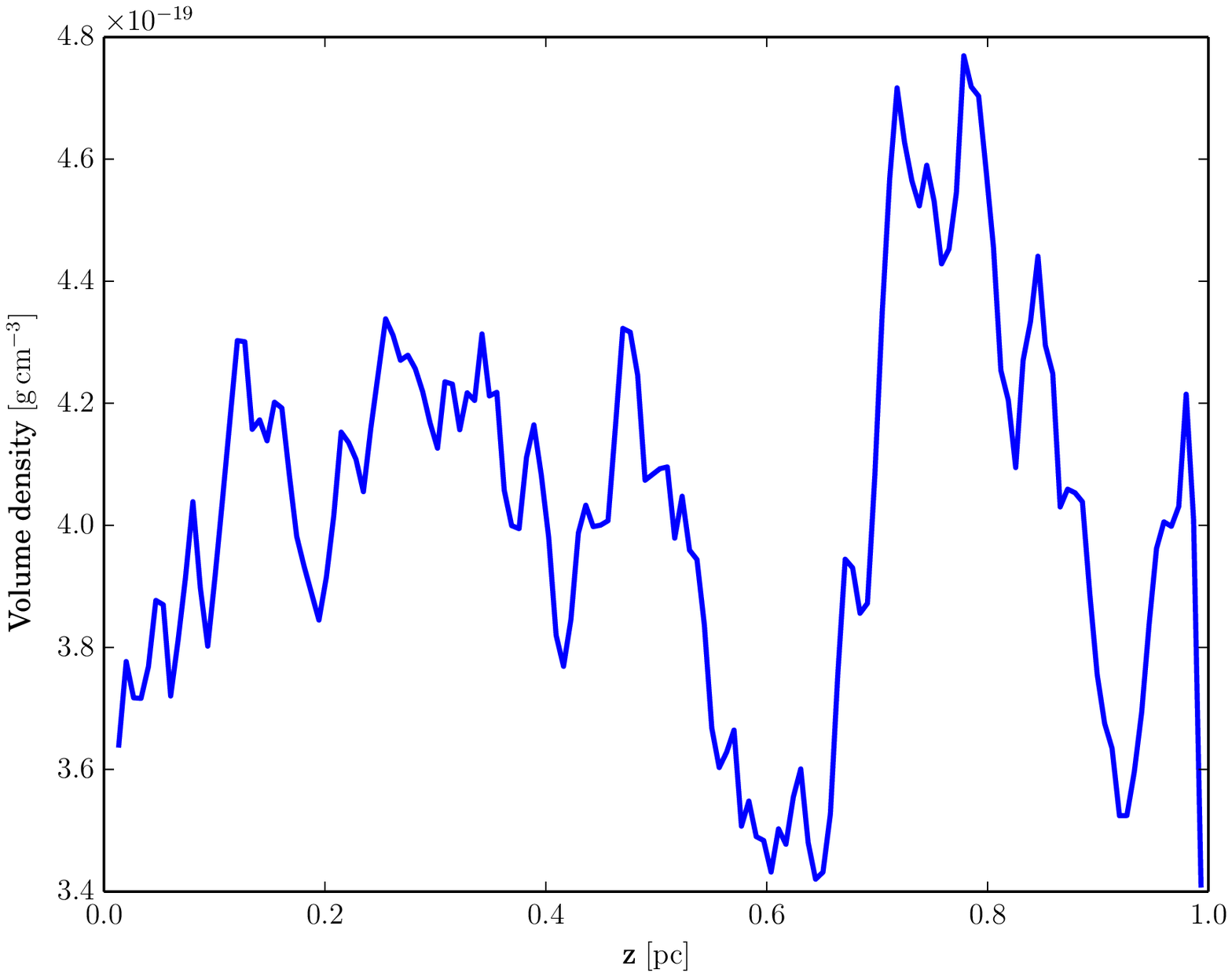}
\includegraphics[width=0.49\linewidth]{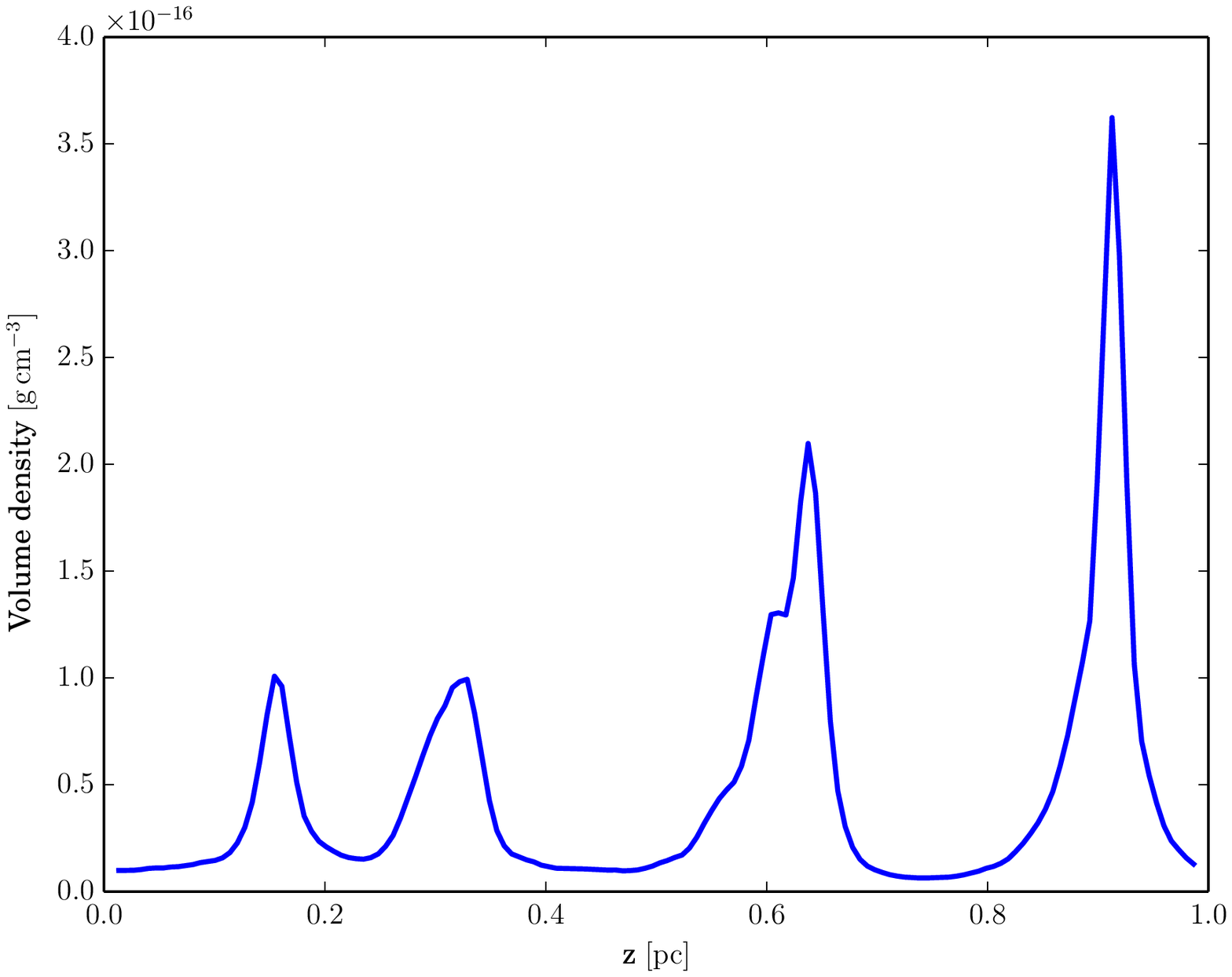}
\caption{The longitudinal density profile for one of the $\alpha = -1.6$ simulations at $t = 0.05 \, \rm Myr$ (left) and $t = 0.56 \, \rm Myr$ (right). The density peaks (cores) at $t = 0.56 \, \rm Myr$, are roughly periodically spaced, having an average separation of $0.24 \, \rm pc$.}
\label{fig::multi}
\end{figure*}

The primary characteristic of the sub-critical phase is that thermal pressure forces are greater than gravitational forces. Thus, during this phase the evolution of the perturbations within the filament is dominated by acoustic oscillations, and standing waves are set up along its length. Figure \ref{fig::den} shows the volume-density profile along the $z$-axis at two different times. The density peaks oscillate with a period $\lambda / a\OO \, \sim 0.50 \rm Myr$. Thus, between 0.05 Myr and 0.30 Myr, a half period has elapsed, and the density peaks and troughs have switched.

It is the presence of these acoustic density oscillations that gives rise to the oscillations seen in the dispersion relation (figure \ref{fig::dis}). The depressed growth rate at certain wavelengths is due to the fact that in these cases the time at which radial collapse begins, $\tau_{_{\rm CRIT}}$, coincides with the point in the oscillation at which the peak of the density perturbation is being dispersed and reformed at a trough (figure \ref{fig::vel}). Thus when the gas making up the density peak becomes self-gravitating it is moving outwards. This has two consequences: the gas must first be decelerated and turned around before the perturbation collapses longitudinally as well as radially, and the amplitude of the density peak is decreased as mass has moved into what once was a trough. These two effects cause the collapse of the perturbation to proceed more slowly once the filament has become super-critical, resulting in a significantly lower growth rate. Conversely, the peaks are caused by the longitudinal motions converging on a density peak at the time of radial collapse, when the perturbation is reaching its peak amplitude.

One can write an equation to predict the positions of the peaks in the dispersion relation by considering when the oscillation period and the time at which the filament becomes supercritical, $\tau_{_{\rm CRIT}}$, are in phase. For the two to be in phase requires

\begin{equation}
\tau_{_{\rm CRIT}} \sim n \tau_{_{\rm HALF \, OSCILLATION}},
\end{equation}
where n is a positive integer and
\begin{equation}
\tau_{_{\rm HALF \, OSCILLATION}} = \frac{\lambda}{2 a\OO}.
\end{equation}
Hence
\begin{equation}
\lambda_{_{\rm {DOM} \, n}} \sim \frac{2 \tau_{_{\rm CRIT}} a\OO}{n}.
\label{eq::peak}
\end{equation}

At $T\OO = 40 \, \rm K$ the isothermal sound speed, $a\OO$, is $0.37 \, \rm{km/s}$, and taking $\tau_{_{\rm CRIT}}$ as 0.45 Myr we expect the dispersion relation to show peaks at $\lambda_{_{\rm PEAK}} = 0.33, 0.17, 0.11, . . .$ corresponding to $n = 1, 2, 3, . . .$. We see the $n \leq 6 $ peaks in figure \ref{fig::dis} where equation \ref{eq::peak} predicts. The dominant wavelength, the one that grows most quickly, is the $n=1$ peak (figure \ref{fig::dis}).

This dominant wavelength, $\lambda_{_{\rm DOM}}$, is defined by a resonance between the timescale, $\tau_{_{\rm CRIT}}$ on which the line density of the filament approaches the critical value for radial collapse, and the timescale on which the longitudinal oscillations of the perturbation complete a half oscillation, $\tau_{_{\rm HALF \, OSCILLATION}}$, so that both motions act to enhance the perturbation - i.e. both radial and longitudinal flows are converging.

From equation \ref{eq::peak} we can see that if the temperature, and hence the sound speed, $a\OO$, are kept constant, and the mass influx per unit length, $\dot{\mu}$, is increased, the dominant wavelength decreases because $\tau_{_{\rm CRIT}}$ has been reduced (see figure \ref{fig::dismu}). Conversely if $\tau_{_{\rm CRIT}}$ is kept constant, and the temperature and sound speed are decreased, the dominant wavelength again decreases, but now because the timescale to complete half a longitudinal oscillation has been increased (see figure \ref{fig::disT}).     

As the oscillations are acoustic in nature and are dependent on the sound speed it is not clear yet if they will survive in the presence of turbulence. However, it has been shown that filaments are decoupled from the supersonic medium surrounding them and appear to only contain sub-sonic motions \citep{HacTaf11}, which may lead to only a small correction to the model presented here.

As noted above, the dispersion relation (figure \ref{fig::dis}) shows that there is an upwards trend in growth rate with increasing wavelength, longer wavelength perturbations grow faster than shorter wavelength perturbations. This can be explained in terms of the Jeans length. The Jeans length is applicable because the perturbations are roughly spherical as they approach instability, as also observed by \citet{InuMiy97}. For a perturbation to be unstable and collapse it's length must be greater than the Jeans length, $\lambda_{J} = a\OO (\pi/G\rho)^{1/2}$. Smaller wavelength perturbations must therefore reach higher densities before they become unstable, and this delays their growth.  

\subsection{Multi-wavelength perturbations}\label{SSEC:DIS2} 

We now apply our analysis to filaments which are seeded with perturbations at multiple wavelengths, in order to investigate whether the dispersion relation continues to hold true. We use initial conditions informed by the work of \citet{Roy15}, who find that the power spectrum of density perturbations in interstellar filaments is well described by a power law with an index of $\alpha = -1.6 \, \pm \, 0.3$. 

We have performed a set of simulations whose initial longitudinal density profile is characterised by such a power law. We perform ten realisations with an index of $\alpha = -1.3$, ten with $\alpha = -1.6$ and ten with $\alpha = -1.9$. So as to compare the results to the dispersion relation (figure \ref{fig::dis}) in section \ref{SEC:RES}, we take $T\OO = 40 \, \rm{K}$, the computational domain is set to $1 \, \rm pc$, $\dot{\mu} = 70 \, \rm{M\Su \, pc^{-1} \, Myr^{-1}}$, the maximum amplitude perturbation has $A=0.2$, the same radial density and velocity profiles are used, and the $k=1$ mode has an amplitude of 0. We set the amplitude of the $k=1$ mode to zero as it corresponds to a wavelength of $1 \, \rm pc$, beyond the range of perturbation wavelengths we considered in the previous section. 

Figure \ref{fig::multi} shows the longitudinal density profile for one of the $\alpha = -1.6$ simulations at $t = 0.05 \, \rm Myr$ and $t = 0.56 \, \rm Myr$ (just before the first sink forms). The density peaks are roughly periodic at $t = 0.56 \, \rm Myr$ despite there being no obvious indication of it in the initial conditions. This suggests that there exists a preferential length scale for fragmentation. 

\begin{figure}
\centering
\includegraphics[width = 0.98\linewidth]{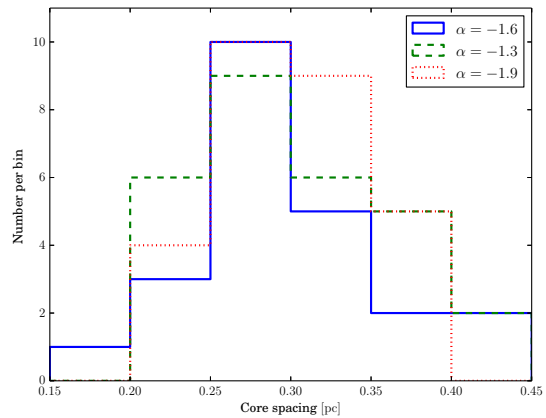}
\caption{Histograms showing the distribution of core spacings in a set of 30 simulations. The blue line shows the core spacings for the ten simulations where $\alpha = -1.6$, the green line shows the core spacings from the ten simulations where $\alpha = -1.3$ and the red line shows the core spacing for the ten simulations where $\alpha = -1.9$. Spacings below $0.05 \, \rm pc$ were removed; in a number of simulations a few cores form very close together, $\sim 0.02 \, \rm pc$, these clusters are then separated by much greater distances. The change in power index does not affect the distribution, all three distributions are strongly peaked at $\sim 0.3 \, \rm pc$. Considering all 30 simulations together, a sample of 114 spacings, the distribution has a mean of 0.296 pc and a standard deviation of 0.070 pc}
\label{fig::hist}
\end{figure}

We determine the core separations for each set of simulations and plot the histograms in figure \ref{fig::hist}. Separation distances below $0.05 \, \rm pc$ have been removed; in a number of simulations a few cores form very close together, with separations $\sim 0.02 \, \rm pc$, and these clusters of cores are then separated by much greater distances.

Changing the index of the power does not appear to affect the distribution of core separations, all three are sharply peaked at $\sim 0.3 \, \rm pc$. As the value of $\alpha$ does not have a strong effect on the distribution we combine the data from all 30 simulations, this leads to a sample of 114 spacings with a mean of 0.296 pc and a standard deviation of 0.070 pc.

The peak in core separations corresponds to the $n=1$ peak in the dispersion relation (figure \ref{fig::dis}); even though there is initially greater power in other modes, it is the wavelength with the fastest growth rate which determines the fragmentation length scale. 

Our simulations suggest that one could estimate a lower age limit for a filament which is fragmenting periodically by measuring the average core separation distance. Equation \ref{eq::peak} can be re-written as,

\begin{equation}
\tau_{_{\rm AGE}} \geq \tau_{_{\rm CRIT}} \simeq \frac{\lambda_{_{\rm CORE}}}{2 a\OO}.
\label{eq::time2}
\end{equation}

This is a lower limit on the filament's age, as we do not know how much time has elapsed between when the cores formed and when the filament is observed. We can also estimate the average accretion rate that the filament experienced during its assembly, $\dot{\mu} = \mu_{_{\rm CRIT}} / \tau_{_{\rm CRIT}} = 2 a\OO \mu_{_{\rm CRIT}} / \lambda_{_{\rm CORE}}$. A filament which has closely spaced cores is likely to have experienced a very high accretion rate and vice versa. 

\citet{TafHac15} find that there exists a preferential core separation of $\sim 0.2 \, \rm pc$ in the sub-filaments making up the L1495/B213 complex in Taurus. Taking a gas temperature of 10 K and assuming solar metallicity the sound speed is $\, 0.19 \, \rm{km/s}$. This results in a minimum age for these sub-filaments of $0.53 \, \rm Myr$. The critical line density at this temperature is $16.7 \, \rm{M\Su \, pc^{-1}}$. Therefore the average accretion rate during the filament's formation was $\sim 32 \, \rm{M\Su \, pc^{-1} \, Myr^{-1}}$, in agreement with the accretion rate inferred observationally by \citet{Pal13}, $27 - 50 \, \rm{M\Su \, pc^{-1} \, Myr^{-1}}$.    

\section{Conclusions}\label{SEC:CON}%

Filament fragmentation is a complex phenomenon. We have shown that the sub-critical accretion phase in a filament's evolution strongly influences the growth of the perturbations which lead to fragmentation. During the sub-critical accretion phase the thermal pressure term dominates over gravity and sets up standing gravo-acoustic oscillations, leading to a series of oscillations in the dispersion relation linking perturbation growth rate and perturbation wavelength. 

The fastest growing mode in an evolving filament is no longer simply linked to the diameter of a filament \citep[as shown to be the case for equilibrium filaments by][]{InuMiy92}. Instead it is dependent on the temperature and the accretion rate onto the filament, because the fastest growing wavelength is the one for which there is a resonance between the timescale on which the filament becomes super-critical and the period of the longitudinal oscillations. We have shown that this dependence holds when filaments are seeded with a multi-wavelength density power spectrum. Moreover, the results are insensitive to the exact power law index in the range of indices ($-1.3 \leq \alpha \leq -1.9$) observed by \citet{Roy15}.

These results allow observers to constrain the age of a filament which is breaking up into regularly spaced fragments \citep[e.g.][]{TafHac15,Beu15} as well as its accretion history.  

\section{Acknowledgments}\label{SEC:ACK}%
SDC gratefully acknowledges the support of a STFC postgraduate studentship. APW gratefully acknowledges the support of a consolidated grant (ST/K00926/1) from the UK Science \& Technology Facilities Council. DAH acknowledges the support of the DFG cluster of excellence ``Origin and Structure of the Universe''. We also thank the anonymous referee for their very useful comments which helped to improve this paper. This work was performed using the computational facilities of the Advanced Research Computing at Cardiff (ARCCA) Division, Cardiff University. 

\bibliographystyle{mn2e}
\bibliography{ref} 

\begin{thebibliography}{25}
\expandafter\ifx\csname natexlab\endcsname\relax\def\natexlab#1{#1}\fi

\bibitem[{{Andr{\'e}} {et~al}\mbox{.}(2010){Andr{\'e}}, {Men'shchikov},
  {Bontemps}, {K{\"o}nyves}, {Motte}, {Schneider}, {Didelon}, {Minier},
  {Saraceno}, {Ward-Thompson}, {di Francesco}, {White}, {Molinari}, {Testi},
  {Abergel}, {Griffin}, {Henning}, {Royer}, {Mer{\'{\i}}n}, {Vavrek}, {Attard},
  {Arzoumanian}, {Wilson}, {Ade}, {Aussel}, {Baluteau}, {Benedettini},
  {Bernard}, {Blommaert}, {Cambr{\'e}sy}, {Cox}, {di Giorgio}, {Hargrave},
  {Hennemann}, {Huang}, {Kirk}, {Krause}, {Launhardt}, {Leeks}, {Le Pennec},
  {Li}, {Martin}, {Maury}, {Olofsson}, {Omont}, {Peretto}, {Pezzuto}, {Prusti},
  {Roussel}, {Russeil}, {Sauvage}, {Sibthorpe}, {Sicilia-Aguilar}, {Spinoglio},
  {Waelkens}, {Woodcraft}, \& {Zavagno}}]{And10}
{Andr{\'e}} P. {et~al.}, 2010, \aap, 518, L102

\bibitem[{{Beuther} {et~al}\mbox{.}(2015){Beuther}, {Ragan}, {Johnston},
  {Henning}, {Hacar}, \& {Kainulainen}}]{Beu15}
{Beuther} H., {Ragan} S.~E., {Johnston} K., {Henning} T., {Hacar} A.,
  {Kainulainen} J.~T., 2015, \aap, 584, A67

\bibitem[{{Burkert} \& {Hartmann}(2004)}]{BurHar04}
{Burkert} A., {Hartmann} L., 2004, \apj, 616, 288

\bibitem[{{Clarke} \& {Whitworth}(2015)}]{ClaWhi15}
{Clarke} S.~D., {Whitworth} A.~P., 2015, \mnras, 449, 1819

\bibitem[{{Fischera} \& {Martin}(2012)}]{FisMar12}
{Fischera} J., {Martin} P.~G., 2012, \aap, 542, A77

\bibitem[{{Freundlich}, {Jog} \& {Combes}(2014){Freundlich}, {Jog}, \&
  {Combes}}]{Fre14}
{Freundlich} J., {Jog} C.~J., {Combes} F., 2014, \aap, 564, A7

\bibitem[{{Hacar} \& {Tafalla}(2011)}]{HacTaf11}
{Hacar} A., {Tafalla} M., 2011, \aap, 533, A34

\bibitem[{{Heitsch}(2013)}]{Hei13a}
{Heitsch} F., 2013, \apj, 769, 115

\bibitem[{{Hubber}, {Walch} \& {Whitworth}(2013){Hubber}, {Walch}, \&
  {Whitworth}}]{Hub13}
{Hubber} D.~A., {Walch} S., {Whitworth} A.~P., 2013, \mnras, 430, 3261

\bibitem[{{Inutsuka}(2001)}]{Inu01}
{Inutsuka} S.-i., 2001, \apjl, 559, L149

\bibitem[{{Inutsuka} \& {Miyama}(1992)}]{InuMiy92}
{Inutsuka} S.-I., {Miyama} S.~M., 1992, \apj, 388, 392

\bibitem[{{Inutsuka} \& {Miyama}(1997)}]{InuMiy97}
{Inutsuka} S.-i., {Miyama} S.~M., 1997, \apj, 480, 681

\bibitem[{{Kirk} {et~al}\mbox{.}(2013){Kirk}, {Myers}, {Bourke}, {Gutermuth},
  {Hedden}, \& {Wilson}}]{Kir13}
{Kirk} H., {Myers} P.~C., {Bourke} T.~L., {Gutermuth} R.~A., {Hedden} A.,
  {Wilson} G.~W., 2013, \apj, 766, 115

\bibitem[{{Lada}, {Alves} \& {Lada}(1999){Lada}, {Alves}, \& {Lada}}]{Lad99}
{Lada} C.~J., {Alves} J., {Lada} E.~A., 1999, \apj, 512, 250

\bibitem[{{Larson}(1985)}]{Lar85}
{Larson} R.~B., 1985, \mnras, 214, 379

\bibitem[{{Morris} \& {Monaghan}(1997)}]{MorMon97}
{Morris} J.~P., {Monaghan} J.~J., 1997, Journal of Computational Physics, 136,
  41

\bibitem[{{Myers}(2009)}]{Mye09}
{Myers} P.~C., 2009, \apj, 700, 1609

\bibitem[{{Ostriker}(1964)}]{Ost64}
{Ostriker} J., 1964, \apj, 140, 1056

\bibitem[{{Palmeirim} {et~al}\mbox{.}(2013){Palmeirim}, {Andr{\'e}}, {Kirk},
  {Ward-Thompson}, {Arzoumanian}, {K{\"o}nyves}, {Didelon}, {Schneider},
  {Benedettini}, {Bontemps}, {Di Francesco}, {Elia}, {Griffin}, {Hennemann},
  {Hill}, {Martin}, {Men'shchikov}, {Molinari}, {Motte}, {Nguyen Luong},
  {Nutter}, {Peretto}, {Pezzuto}, {Roy}, {Rygl}, {Spinoglio}, \&
  {White}}]{Pal13}
{Palmeirim} P. {et~al.}, 2013, \aap, 550, A38

\bibitem[{{Pon}, {Johnstone} \& {Heitsch}(2011){Pon}, {Johnstone}, \&
  {Heitsch}}]{Pon11}
{Pon} A., {Johnstone} D., {Heitsch} F., 2011, \apj, 740, 88

\bibitem[{{Price}(2007)}]{Pri07}
{Price} D.~J., 2007, \pasa, 24, 159

\bibitem[{{Price} \& {Monaghan}(2004)}]{PriMon04}
{Price} D.~J., {Monaghan} J.~J., 2004, \mnras, 348, 139

\bibitem[{{Roy} {et~al}\mbox{.}(2015){Roy}, {Andre'}, {Arzoumanian}, {Peretto},
  {Palmeirim}, {Konyves}, {Schneider}, {Benedettini}, {Di Francesco}, {Elia},
  {Hill}, {Ladjelate}, {Louvet}, {Motte}, {Pezzuto}, {Schisano}, {Shimajiri},
  {Spinoglio}, {Ward-Thompson}, \& {White}}]{Roy15}
{Roy} A. {et~al.}, 2015, ArXiv e-prints

\bibitem[{{Schneider} \& {Elmegreen}(1979)}]{SchElm79}
{Schneider} S., {Elmegreen} B.~G., 1979, \apjs, 41, 87

\bibitem[{{Tafalla} \& {Hacar}(2015)}]{TafHac15}
{Tafalla} M., {Hacar} A., 2015, \aap, 574, A104

\end{thebibliography}

\label{lastpage}

\end{document}